\documentclass{ws-mpla}
\usepackage{epsf,epsfig,subfigure,axodraw,graphicx,amsmath,amssymb}

\def\EE{E$_8\times$E$_8^\prime$}
\def\Eo{E$_8$}

\def\Sp{${\cal S}$}
\def\Hvec{${\cal V}$}

\begin{document}

\markboth{J. E. Kim} {SSM from Z$_{12-I}$}

\catchline{}{}{}{}{}

\title{ ${\bf Z}_{12-I}$
Orbifold Compactification toward SUSY Standard Model }

\author{\footnotesize Jihn E. Kim}

\address{Department of Physics and Astronomy, Seoul National
University, \\
Seoul 151-747, Korea\\
jekim@phyp.snu.ac.kr}

\maketitle


\begin{abstract}
We explain the orbifold compactification in string models and
present a ${\bf Z}_{12-I}$ orbifold compactification toward
supersymmetric standard models. We also point out an effective
$R$-parity from this string construction. The VEVs of gauge singlets
are chosen such that phenomenological constraints are satisfied.

\end{abstract}

\ccode{PACS Nos.: 14.80.Mz, 11.25.Mj, 11.25.Wx, 11.30.Fs}

\section{Introduction}

The standard model (SM) with 45(+3) chiral fields is really
remarkable. The big question in particle physics is, $\lq\lq$How
does this SM arise?" In late 1980s, there were attempts for
standard-like models (which has three families with the gauge group
SU(3)$\times$SU(2)$\times$U(1)$^n$) \cite{iknq} from the orbifold
compactification of heterotic string. Recent attempts have been more
ambitious but it is fair to say that a model free of any
phenomenological problems has not appeared yet even though partially
attractive ones have been proposed in trinification, Pati-Salam, or
just SM\cite{SMparttial}. So, searches for good supersymmetric
(SUSY) standard models (SSM) are going on now vigorously. Even some
string models are suggested as roots for explaining the PVLAS data
\cite{PVLAS}. In this talk, I follow the compactification route
through orbifolds. Orbifolds are manifolds modded by discrete
actions. A nice feature of the orbifold compactification is that it
is basically a geometric one.

For an SSM, we may obtain it either directly from compactification
or through an intermediate step of SUSY GUT. In ${\bf Z}_{12-I}$, we
have constructed both kinds \cite{KKK,KKflip}, and here we focus on
the direct construction.

In the orbifold construction, there has been the adjoint difficulty
that at the Kac-Moody level $k=1$ there does not appear an adjoint
matter representation. Thus, GUTs SU(5), SO(10), and E$_6$ are not
good candidates toward SSMs simply because the Higgs mechanism for
breaking the GUT group is not present. This prefers GUTs with factor
groups such as SU(3)$^3$, SU(4)$\times$SU(2)$\times$SU(2) and
SU(5)$\times$U(1). The trinification SU(3)$^3$ is possible only in
${\bf Z}_3$ which is nice toward achieving \cite{Kim03}
$\sin^2\theta_W=\frac38$. In addition to the $\sin^2\theta_W$
problem, there are several problems to be explained in those models,
\begin{itemize}
\item Approximate $R$-parity for proton longevity,
\item Exotics problem,
\item Vectorlike pairs problem,
\item Successful fit to quark and lepton masses and mixing angles,
\item Strong CP problem \cite{staxion,QCDaxZ12}, etc.
\end{itemize}
Among all these problems, the most difficult and urgent one to
overcome is the $R$-parity problem. One of the nice features of
SO(10) GUT is said to be that it has a scheme to introduce the
$R$-parity. Noting that SO(10) has both spinor (${\cal S}$) and
vector (${\cal V}$) representations, one can assign $R=-1$ for
${\cal S}$ and $R=+1$ for ${\cal V}$ and then tree level couplings
respect the $R$-parity. However, this is true only when gauge
singlets are not introduced. The gauge singlets present in string
compactification may behave like a spinor or a vector and the above
simple argument of SO(10) GUT is not applicable to SSMs from string
compactification. Thus, the $R$-parity consideration is most
important. Only, approximate $R$-parity is obtained in string models
so far \cite{Rparity,KKK}. Previous SSMs from string have not
obtained the $R$-parity problem properly.
\section{Strings on Orbifolds}

Orbifolds are manifolded moded by discrete actions. The simplest
example is $S_1/{\bf Z}_2$. In string compactifications, six
internal spaces are usually divided into three two tori $T^2\otimes
T^2\otimes T^2$. Each $T^2$ can be moded out be a discrete action.
The totality of each discrete action is given a name ${\bf Z}_N$
orbifold.

\begin{figure}[h]
\begin{picture}(400,100)(0,0)
\CArc(180,50)(50,0,360) {\SetWidth{1.6} \CArc(180,50)(50,270,90)}
 \Text(180,0)[c]{$\bullet$}\Text(180,100)[c]{$\bullet$}
 \LongArrow(170,50)(132,50)\Text(180,50)[c]{${\bf Z}_2$}
 \LongArrow(190,50)(228,50)
 \LongArrow(170,80)(142,80)\Text(180,80)[c]{${\bf Z}_2$}
 \LongArrow(190,80)(218,80)
 \LongArrow(170,20)(142,20)\Text(180,20)[c]{${\bf Z}_2$}
 \LongArrow(190,20)(218,20)
\end{picture}
\caption{The simplest orbifold  $S_1/{\bf Z}_2$.}\label{fig:S1}
\end{figure}
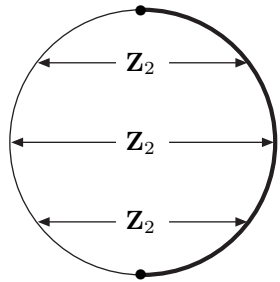

\begin{figure}[h]
\begin{picture}(400,110)(0,0)
\LongArrow(80,0)(80,110) \LongArrow(80,0)(260,0)
\GBox(80,0)(160,90){0.8} \Line(160,90)(240,90) \Line(240,90)(240,0)
\Line(160,0)(240,0)
 \Line(120,95)(110,85) \Line(123,95)(113,85)
 \Line(120,5)(110,-5) \Line(123,5)(113,-5)
 \Line(85,25)(75,15) \Line(85,70)(75,60)
\Line(165,25)(155,15) \Line(165,70)(155,60)
 \Line(165,27)(155,17) \Line(165,72)(155,62)
  \Line(165,23)(155,13) \Line(165,68)(155,58)

\Text(80,45)[c]{$\bullet$}\Text(160,45)[c]{$\bullet$}
\Text(80,0)[c]{$\bullet$}\Text(160,0)[c]{$\bullet$}
\end{picture} \caption{The $T^2/ {\bf Z}_2$ orbifold with
two different length scales. The fundamental region is grey-colored
which becomes a pillow after the identifications. The fixed points
are at the boundary.}\label{fig:Z2}
\end{figure}
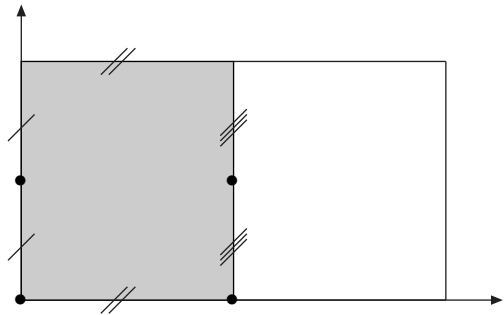

The simplest orbifold is moding $T^1$ by  ${\bf Z}_2$, identifying
two points connected by  ${\bf Z}_2$ actions. The points which stays
at the same point under the ${\bf Z}_2$ action is called fixed
points. Pictorially, we show this $S_1/{\bf Z}_2$ orbifold in Fig.
\ref{fig:S1} with fixed points located at {\it the boundary of the
fundamental region} shown as the thick arc.
Another frequently discussed orbifold is $T^2/ {\bf Z}_2$ shown in
Fig. \ref{fig:Z2}, which has four fixed points.
For a ${\bf Z}_N$ orbifold, satisfying $\theta^N=1$ for the
consistency of the worldsheet spinors, we have
\begin{equation}
N\sum_i\phi_i={\rm even\ integer}.
\end{equation}

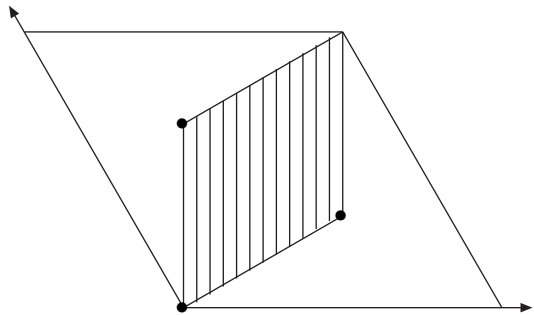
\begin{figure}[h]
\begin{picture}(400,110)(0,0)
\LongArrow(140,0)(270,0) \LongArrow(140,0)(75,112.6)
\Line(80,104)(200,104)
 \Line(260,0)(200,104)
 \Line(140,0)(140,69.33) \Line(140,69.33)(200,104)
 \Line(200,104)(200,34.67)\Line(200,34.67)(140,0)
 \Text(200,34.67)[c]{$\bullet$} \Text(140,0)[c]{$\bullet$}
  \Text(140,69.33)[c]{$\bullet$}
  \Line(195,102)(195,32.67)\Line(190,99)(190,29.67)
  \Line(185,96)(185,26) \Line(180,93)(180,23) \Line(175,90)(175,20)
  \Line(170,87)(170,17) \Line(165,84)(165,14) \Line(160,81)(160,11)
  \Line(155,78)(155,8) \Line(150,75)(150,5) \Line(145,72)(145,2)
\end{picture}
\caption{The $T^2/ {\bf Z}_3$ orbifold with 120$^o$ rotation
actions. The fundamental region is shown as slashed
vertically.}\label{fig:Z3}
\end{figure}

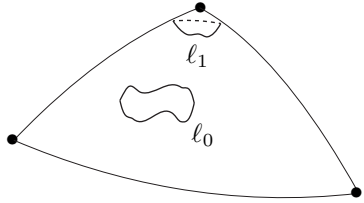
\begin{figure}[h]
\begin{center}
\begin{picture}(400,90)(0,0)

 {\SetWidth{0.3}
 \Curve{(110,30)(120,40)(180,80)} \Text(110,30)[c]{$\bullet$}
  \Text(181,80)[c]{$\bullet$}
 \Curve{(180,80)(228,30)(240,10)} \Curve{(110,30)(170,12)(240,10)}
 \Text(240,10)[c]{$\bullet$}
 }
    \Curve{(170,75)(175,70)(180,70)(183,69)(188,74)}
    \DashCurve{(170,75)(180.5,75)(188,74)}{1.5}
    \Text(180,62)[c]{$\ell_1$}
 \Curve{(150,45)(150.5,46)(153,50)(158,48)(173,50)(177.5,46)(178,44)}
 \Curve{(150,45)(150.5,43)(155,38)(170,40)(173,37)(177.5,42)(178,44)}
     \Text(182,32)[c]{$\ell_0$}

\end{picture}
\caption{The untwisted string $\ell_0$ and twisted string $\ell_1$
are shown.}\label{fig:UT}
\end{center}
\end{figure}

The most widely discussed and relevant one for us in this talk is
 $T^2/ {\bf Z}_3$ orbifold shown in Fig. \ref{fig:Z3}.
For the coordinates of three two-tori, we use the complexified
coordinates,\footnote{For a more concrete discussion, see
\cite{Orbbook}.}
\begin{equation}
z\equiv (z_1,z_2,z_3)\sim \theta \cdot z=(e^{2\pi i
\phi_1}z_1,e^{2\pi i \phi_2}z_2,e^{2\pi i \phi_3}z_3)
\end{equation}
where $\phi_i$ denote the rotation angle in the $i^{\rm th}$ torus.
Since one $T^2/ {\bf Z}_3$ orbifold has three fixed points, the
$(T^2/ {\bf Z}_3)^3$ orbifold has 27 fixed points. There is only one
way to write the {\it twist vector} $\phi$,\footnote{For ${\bf
Z}_{6}, {\bf Z}_{8}$, and ${\bf Z}_{12}$, there are two kinds of
twists denoted by  ${\bf Z}_{6-I},  {\bf Z}_{6-II}, {\bf
Z}_{8-I},{\bf Z}_{8-II},{\bf Z}_{12-I}$, and ${\bf Z}_{12-II}$.}
\begin{equation}
\phi=\textstyle (\frac23,\frac13,\frac13).
\end{equation}
In quantum mechanics, any symmetry action is embedded in the quantum
mechanical group space, here in particular in the \EE\ space. The
simplest example called the standard embedding is the following
embedding \cite{DHVW},
$$
V=\textstyle(\frac23~\frac13~\frac13~0~0~0~0~0)(0^8)'
$$
which breaks \Eo\ down to E$_6\times$SU(3). Particle states are
classified by untwisted and twisted sectors. Particles in the
untwisted (U) sector are the closed strings in the torus. Particles
in the twisted (T) sector are the closed strings only after the
discrete action is taken into account. These U and T sector strings
for a ${\bf Z}_3$ torus are shown in Fig. \ref{fig:UT}.
The T strings are located at fixed points as shown in Fig.
\ref{fig:UT}. There is additional degrees of freedom to embed the
action in the group space \cite{inq}, called the Wilson lines.
Wilson lines are embedded in the group space, for example as
$(0~0~0~\frac23~\frac13~\frac13~0~0)(0^8)'$. There are consistency
conditions for the shift vector $V$ and the Wilson lines to satisfy
for the modular invariance. The Wilson lines can break the group
further. Without a Wilson line in the $i^{\rm th}$ torus, three
fixed points of the $i^{\rm th}$ torus is not distinguishable and
theory must respect an $S_3$ permutation symmetry \cite{S3Z3}. The
three fixed points of the $i^{\rm th}$ torus are distinguished by a
Wilson line in the $i^{\rm th}$ torus, i.e. by $V+a_i, V-a_i$, and
$V$.

The massless U strings satisfy just the modular invariance
conditions with $V$ and Wilson lines $a_i$. If $P$ is a weight in
the group space the massless U sector strings must satisfy, for
${\bf Z}_3$ for example,
\begin{align}
\begin{array}{l}
P\cdot V=\textstyle 0,\pm\frac13\ {\rm mod\ integer}\\
P\cdot a_i=0,\ {\rm mod\ integer}
\end{array}\label{ModCond}
\end{align}
Since Wilson lines distinguish fixed points, the U sector strings
are not affected by their presence except the modular invariance
condition (\ref{ModCond}). Considering the untwisted sector vacuum
energy, there is a masslessness spectrum condition also \cite{DHVW}.

The T sector strings are made closed by the orbifolding action.
Thus, the masslessness spectrum condition in the T sector is
different from that of the U sector \cite{DHVW}. The modular
invariance of the theory is given by the required form for $V$ and
$a_i$, and in the T sector distinguishing the fixed points we do not
require any more. But in the T sector where Wilson lines do not
distinguish fixed points, we require the condition similar to
(\ref{ModCond}) of the U sector.

The method of obtaining orbifold models is explicitly illustrated
below with a ${\bf Z}_{12-I}$ twist and shift vectors $V$ and
$a_3=a_4$.

\section{Model}

The ${\bf Z}_{12-I}$ twist is
\begin{equation}
\phi=\textstyle(\frac{5}{12},\frac{4}{12},\frac{1}{12})\label{twist}
\end{equation}
and we take the following shift vector $V$ and Wilson line
\cite{KKK}
\begin{align}
&V=\textstyle \frac{1}{12}(3~3~3~3~3~5~5~1)(3~9~0^6)'\label{shiftV}\\
&a_3=a_4=\textstyle \frac13(2~2~2~-2~-2~2~0~2)(0~2~2~0^5)'\label{shifta}\\
&a_1=a_2=a_5=a_6=0.
\end{align}
From Eq. (\ref{twist}), we note that (12)- and (56)-tori are truly
${\bf Z}_{12}$ moding while the (34)-torus is  ${\bf Z}_{3}$ moding.
Therefore, Wilson lines distinguishing fixed points are applicable
only to the (34)-torus and that must satisfy a  ${\bf Z}_{3}$ shift
as shown in (\ref{shifta}). Thus, for ${\bf Z}_{12-I}$ the Wilson
line part is very simple, i.e. it is just distinguishing three fixed
points of the (34)-torus. In this sense,  ${\bf Z}_{12-I}$ is a very
simple model. For the gauge symmetry breaking, already there are
much more possibilities of breaking even without Wilson lines in
${\bf Z}_{12-I}$ because many integers can be assigned in the
numerator of $\frac{n}{12}$. In this sense,  ${\bf Z}_{12-I}$ is
very simple and also has a simple geometrical interpretation.

The gauge group is obtained by counting massless vector multiplets
which appear in the untwisted sector,
\begin{align}
\begin{array}{l}
P\cdot V=\textstyle 0,\ {\rm mod\ integer}\\
P\cdot a_3=0,\ {\rm mod\ integer}
\end{array}\label{Z12Gauge}
\end{align}
If we consider $V^\pm=V\pm a_3$ as the twisted sector, a similar
condition with $V^\pm$ instead of $V$ can give gauge groups in the
corresponding twisted sectors. In fact, the gauge group obtained
from (\ref{Z12Gauge}) is the intersection of gauge groups of $V$ and
$V^\pm$ which is automatically incoded by the second condition of
(\ref{Z12Gauge}). This gauge group is the one obtained from the U
sector vector multiplets. From (\ref{Z12Gauge}), we obtain the
following gauge group,
\begin{equation}
SU(3)\times SU(2)\times U(1)_Y\times U(1)^4\times [SO(10)\times
U(1)^3]'.
\end{equation}
Embedding of the electroweak hypercharge $Y$ is possible for two
cases for which different weak mixing angles are obtained,
\begin{align}
&{\rm Model\ E}:\ Y=\textstyle(\frac13~\frac13~\frac13~\frac{-1}2
~\frac{-1}2~0~0~0)(0^8)',\quad \sin^2\theta_W=\frac38 \\
&{\rm Model\ S}:\ Y=\textstyle(\frac13~\frac13~\frac13~\frac{-1}2
~\frac{-1}2~0~0~0)(0~0~1~0^5)',\quad \sin^2\theta_W=\frac3{14}.
\end{align}
For Model E, we obtain vectorlike exotics while in Model S there
does not appear any exotics.\footnote{Note added: Another exotics
free model \cite{GMSBst} also has the hypercharge invaded by the
hidden sector. At present, I do not know any exotics free SSM
without the invasion of the electroweak hypercharge by the hidden
sector.} Below, we discuss Model E in detail.

Matter particles are obtained from the U sector and the T sector.
For the U sector, the modular invariance condition for chiral matter
is $P\cdot V=$ one of the entries of $\phi$ given in Eq.
(\ref{twist}), denoted as $U_1$ if it comes from matching with the
first entry, etc. In fact, the CTP conjugates also appear in one of
these. The same chirality set is
$\{\frac{1}{12},\frac{4}{12},\frac{7}{12}\}$.

The fundamental twist $\phi$ is the defining one and called the
$T_1$ sector. If the twist is $\phi$, then any integer ($n$)
multiple of $\phi$ must be considered also. Thus, we consider $T_n$
twist sectors up to $n=6$.\footnote{For $6<n\le 11$, they provide
the CTP conjugates of those appearing in $1\le n< 6$ except in $T_3$
and $T_9$. $T_3$ and $T_6$ states can contain CTP conjugates also.}
Since $a_3$ is a ${\bf Z}_{3}$ shift 3, 6, 9 multiples of twist do
not have any Wilson line. The other twist sectors, however, can have
Wilson lines. Thus, the twist sectors are denoted as
$$
T_1^0, T_1^\pm, T_2^0, T_2^\pm, T_3, T_4^0, T_4^\pm, T_5^0, T_5^\pm,
T_6.
$$
Twisted sector massless condition is given by considering the
corresponding twisted sector vacuum energy, and there is a
well-defined way to calculate their chiralities \cite{Orbbook}.
Since $T_3,T_6,T_9$ sectors do not involve Wilson lines, in addition
they must satisfy in a sense an untwisted sector-like modular
invariance condition similar to (\ref{Z12Gauge}), generalized to
\begin{equation}
(P+kV)\cdot a_3=0\ {\rm mod}\ {\bf Z},\quad {\rm for}\ k=0,3,6,9.
\end{equation}
The calculation of spectrum is not enough. We have to find out the
chirality by considering the right movers. Also, there can be some
linear combinations of localized fields whose multiplicities must be
calculated. After all these considerations, we obtain a complete
massless spectrum. For the method, refer to
\cite{Orbbook,KKflip,KKK}. The SM particles are listed in Table
\ref{tb:SM}. Summarizing the standard charge partcles except extra
neutral singlets
\begin{align}
\begin{array}{l}
U:\ Q(U_1), L(U_1), u^c(U_3), d^c(U_3), \nu^c(U_3), e^c(U_3),
H_u(U_2), H_d(U_2)\\
T_4^0:\ 2\{Q, L, u^c, d^c, \nu^c, e^c, \overline{D}, D, H_u, H_d\},\
\overline{D}, H_d\\
T_6:\ 3\{ \overline{D}, D\},\ 2\{H_u, H_d\}\\
T_3:\  \overline{D}, H_d,\ 3\cdot{\bf 10}'\\
T_9:\ 2D, 2H_u
\end{array}
\end{align}
Note that we can assign the 3rd family in the U sector and the first
and second families in the $T_4^0$ sector. It has been shown that
Yukawa couplings with appropriate neutral singlets can make
vectorlike pairs massive.
\begin{table}[t]
\tbl{The SM particles.}
 {\begin{tabular}{@{}cccc@{}} \toprule \hline  Visible
states & SM notation& $\Gamma$ & $\Gamma'$
\\
\hline
 $(\underline{++-};\underline{+-};+++)(0^8)'$ & $Q(U_1)$ &--1 & +1
\\
$(\underline{+--};--;+++)(0^8)'$ & $d^c(U_3)$ & --1 & +1
\\
$(\underline{+--};++;+--)(0^8)'$ & $u^c(U_3)$ &--1 & $-3$
\\
$(---;\underline{+-};+--)(0^8)'$ & $L(U_1)$ & --1 & $-3$
\\
$(+++;--;-+-)(0^8)'$ & $e^c(U_3)$ & +5 & +5
\\
$({+++;++};+++)(0^8)'$ & $\nu^c(U_3)$ &--1 & +1
\\
$(0~0~0;\underline{-1~0};-1~0~0)(0^8)'$ & $H_u(U_2)$ & +2 & +2
\\
$(0~0~0;\underline{1~0};0~0~1)(0^8)'$ & $H_d(U_2)$ & --4 & $-2$
\\
 $(\underline{++-};\underline{+-};\frac16~\frac16~
 \frac{-1}{6})(0^8)'$ & $2\cdot Q(T_4^0)$ &+1 & +1
 \\
$(\underline{+--};{--};\frac16~\frac16~
 \frac{-1}{6})(0^8)'$ & $2\cdot d^c(T_4^0)$ &+1 & +1
\\
 $(\underline{+--};{++};\frac16~\frac16~
 \frac{-1}{6})(0^8)'$ & $2\cdot u^c(T_4^0)$ & $-3$ & $-3$
\\
$({---};\underline{+-};\frac16~\frac16~
 \frac{-1}{6})(0^8)'$ & $2\cdot L(T_4^0)$ & $-3$ & $-3$
\\
$({+++};{--};\frac16~\frac16~
 \frac{-1}{6})(0^8)'$ & $2\cdot e^c(T_4^0)$ &+5 & +5
\\
$({+++};{++};\frac16~\frac16~
 \frac{-1}{6})(0^8)'$ & $2\cdot \nu^c(T_4^0)$ &+1 & +1
\\
$(\underline{1,0,0};0~0;\frac{-1}{3}~\frac{-1}{3}~\frac{1}{3})(0^8)'$
& $3\cdot \overline{D}_{1/3}(T_4^0)$ & $\fbox{+2}$ & $\fbox{+2}$\\
[0.1em]
$(\underline{-1,0,0};0~0;\frac{-1}{3}~\frac{-1}{3}~\frac{1}{3})(0^8)'$
& $2\cdot {D}_{-1/3}(T_4^0)$ &$\fbox{$-2$}$ & $\fbox{$-2$}$\\
$({0,0,0};\underline{-1~0};\frac{-1}{3}~
\frac{-1}{3}~\frac{1}{3})(0^8)'$ & $2\cdot H_u(T_4^0)$ &+2 & +2\\
$({0,0,0};\underline{1~0};\frac{-1}{3}~\frac{-1}{3}~\frac{1}{3})(0^8)'$
& $3\cdot H_d(T_4^0)$ &--2 & $-2$\\
$(\underline{1,0,0};{0~0};0^3)(\frac{-1}{2}~\frac12~0;0^5)'$ &
$3\cdot \overline{D}_{1/3}(T_6)$ & $\fbox{+2}$ & $\fbox{$+2$}$
\\ [0.1em]
$(\underline{-1,0,0};{0~0};0^3)(\frac{1}{2}~\frac{-1}{2}~0;0^5)'$
& $3\cdot {D}_{-1/3}(T_6)$ & $\fbox{$-2$}$ & $\fbox{$-2$}$\\
 $({0,0,0};\underline{-1~0};0^3)(\frac{-1}{2}~\frac12~0;0^5)'$ &
$2\cdot H_u(T_6)$ &+2 & +2\\
 $({0,0,0};\underline{1~0};0^3)(\frac{1}{2}~\frac{-1}{2}~0;0^5)'$ &
$2\cdot H_d(T_6)$ &$-2$ & $-2$\\

 $(\underline{\frac34\frac{-1}{4}\frac{-1}{4}};{\frac{-1}{4}
  \frac{-1}{4}};\frac{1}{4}\frac{1}{4}\frac{1}{4})
  (\frac{3}{4}\frac{1}{4}~0;0^5)'$ &
$\overline{D}_{1/3}(T_3)$ &${1}$ & \fbox{+2}\\
  $(\underline{\frac{-3}{4}\frac{1}{4}\frac{1}{4}};{\frac{1}{4}
  \frac{1}{4}};\frac{-1}{4}\frac{-1}{4}\frac{-1}{4})
  (\frac{-3}{4}\frac{-1}{4}~0;0^5)'$ &
$2\cdot {D}_{-1/3}(T_9)$ &${-1}$ & \fbox{$-2$}\\
  $({\frac{1}{4}\frac{1}{4}\frac{1}{4}};\underline{\frac{-3}{4}
  \frac{1}{4}};\frac{-1}{4}\frac{-1}{4}\frac{-1}{4})
  (\frac{1}{4}\frac{3}{4}~0;0^5)'$ &
$2\cdot H_u(T_9)$ &${+4}$ & +3\\
  $({\frac{-1}{4}\frac{-1}{4}\frac{-1}{4}};\underline{\frac{3}{4}
  \frac{-1}{4}};\frac{1}{4}\frac{1}{4}\frac{1}{4})
  (\frac{-1}{4}\frac{-3}{4}~0;0^5)'$ &
$H_d(T_3)$ &${-4}$ & $-3$
\\[0.2em]
\hline
\end{tabular}}\label{tb:SM}
\end{table}

The study of exotic particles is very tricky, but in our model these
are known to be vectorlike and made massive by choosing appropriate
VEVs of gauge singlets \cite{KKK}.

There exist eight U(1) gauge symmetries whose charges are denoted as
\begin{align}
&Y=\textstyle
(\frac13~\frac13~\frac13~\frac{-1}2~\frac{-1}2~0^3)(0^8)'\\
&B-L=\textstyle
(\frac23~\frac23~\frac23~0~0~0^3)(0^8)'\label{BmL}\\
&Q_1=(0^5~2~0~0)(0^8)'\label{Q1}\\
&Q_2=(0^5~0~2~0)(0^8)'\label{Q2}\\
&Q_3=(0^5~0~0~2)(0^8)'\label{Q3}\\
&Q_4=(0^8)(2~0~0~0^5)'\nonumber\\
&Q_5=(0^8)(0~2~0~0^5)'\nonumber\\
&Q_6=(0^8)(0~0~2~0^5)'
\end{align}
One linear combination of the above charges is the U(1)$_X$ charge
of the flipped SU(5),
\begin{equation}
X=\textstyle (2~2~2~2~2~0~0~0)(0^8)'
\end{equation}
There exists an anomalous U(1)$_{\rm anom}$ with
\begin{equation}
Q_{\rm anom}=\textstyle Q_1+Q_2+Q_3+Q_4-Q_5+6X.
\end{equation}
Toward embedding a ${\bf Z}_2$ matter parity $P$ in an anomaly free
U(1), we choose U(1)$_\Gamma$ where
\begin{equation}
{\Gamma}=\textstyle
X+\frac14(Q_4+Q_5)-(Q_2+Q_3)+6(B-L).\label{U1Gamma}
\end{equation}

\section{Phenomenology}

Except the gauge interactions, phenomenology results from Yukawa
couplings including the nonrenormalizable terms. The Yukawa coupling
structure respects
\begin{itemize}
\item Gauge symmetries,
\item Lorentz symmetry, in particular from the internal coordinates the
$H$-momentum conservation,
\item The modular invariance conditions for the $T_k^{m_f}$ sector
\begin{align}
&\sum_z k(z)=0\ {\rm mod\ }12\\
&\sum_z [km_f](z)=0\ {\rm mod\ }3
\end{align}
\item The modular invariance requires the sum of $H$-momenta being
$(-1, 1, 1)$ mod (12, 3, 12).\footnote{An extensive discussion is
given in \cite{Orbbook}.}
\end{itemize}
For ${\bf Z}_{12-I}$, the $H$-momenta are given by
\begin{align}
&U_1: (-1,0,0),\quad U_2: (0,1,0),\quad U_3:
(0,0,1),\nonumber\\
&\textstyle T_1:(\frac{-7}{12},\frac{4}{12},\frac1{12}),\quad
 T_2:(\frac{-1}{6},\frac46,\frac16),\quad T_3:
 (\frac{-3}{4},0,\frac{1}{4}),\nonumber\\
&\textstyle
 T_4:(\frac{-1}{3},\frac13,\frac13),\quad
\left\{T_5:(\frac{1}{12},\frac{-4}{12},\frac{-7}{12})\right\}, \quad
T_6:(\frac{-1}{2},0,\frac12),\\
&\textstyle T_7:(\frac{-1}{12},\frac{4}{12},\frac{7}{12}),\quad
T_9:(\frac{-1}{4},0,\frac{3}{4}).  \nonumber
\end{align}
For example, $T_2T_4T_6$ has the $H$-momentum (--1, 1, 1); thus the
coupling is allowed if the other conditions are satisfied. The 3rd
family quark masses arise from $Q(U_1)u^c(U_3)H_u(U_2)$ and
$Q(U_1)d^c(U_3)H_d(U_2)$ which are cubic. But we need much more
couplings to make this model phenomenologically successful. Since
there are O(100) chiral fields, a computer search may be necessary.
We have shown \cite{KKK} that if neutral singlets are allowed to get
GUT scale VEVs then all the needed phenomenology can be met, in
particular vectorlike exotics and non-exotic vectorlike pairs obtain
large masses. Also, supersymmetric conditions the $F$-flatness and
$D$-flatness have to be checked in a specific model for given
singlet VEVs \cite{KKK}. But the $R$-parity needs a special
treatment on which we will comment shortly.

\begin{figure}[pt]
\centerline{\psfig{file=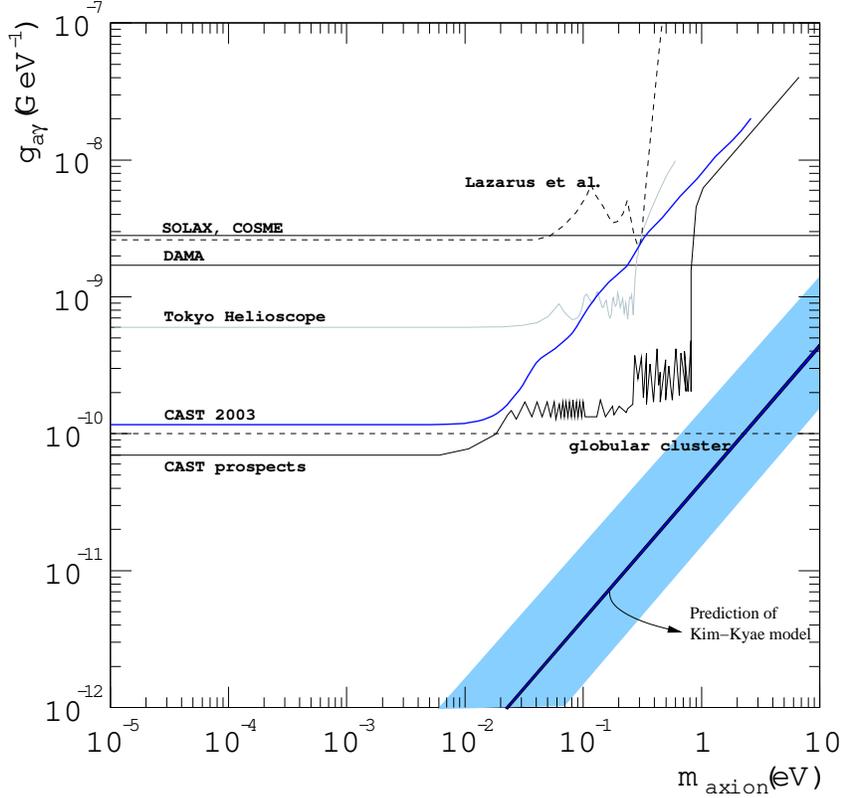,width=12cm}} \vspace*{8pt}
\caption{The CAST 2003 data compared to the  ${\bf Z}_{12}$ model
given by the thick line. The band is the 20 \% theoretical error of
Ref. [15]. } \label{axionexp}
\end{figure}

Before discussing the $R$-parity, we point out that now it is
possible to study an approximate global symmetry. In fact, in the
${\bf Z}_{12-I}$ flipped SU(5) model \cite{KKflip}, we studied an
approximate Peccei-Quinn symmetry \cite{QCDaxZ12}. We studied
U(1)$_{\rm anom}\times$U(1)$_{\rm global}$ where U(1)$_{\rm global}$
is an approximate one. We find that the QCD axion is possible and
obtained the axion-photon-photon coupling for the first time from
string compactification,
\begin{equation}
c_{a\gamma\gamma}=\overline{c}_{a\gamma\gamma}-1.93\simeq -0.26
\end{equation}
where --1.93 appears from the QCD chiral symmetry breaking. However,
the decay constant turns out to be at the GUT scale and hence it is
very difficult to observe the very light axion from the cavity
experiments even though we invoke the anthropic principle for the
misalignment problem.  In Fig. \ref{axionexp}, the above
axion-photon-photon coupling strength is compared to the recent
solar axion search from CAST \cite{Zioutas:2004hi}.

This kind of study on approximate symmetries can be done in a
specific model. In particular, the $R$-parity must be studied in a
specific model as performed in the flipped SU(5) \cite{Rparity}. A
probable failure in obtaining an $R$-parity is that one needs many
singlet VEVs for other phenomenological reasons.

\section{Effective $R$-parity}

The approximate $R$-parity in Ref. [8] was obtained by studying
Yukawa couplings up to dimension 7. But a better way is to embed the
$R$-parity or matter parity\footnote{Strictly speaking, we work with
the matter parity since gauginos are not considered, but this matter
parity can be properly extended to become an $R$-parity. So we use
both words without distinction.} in an anomaly free U(1) gauge
group. A parity is a ${\bf Z}_2$ operation. Some VEVs of the U(1)
charge carrying fields can break U(1) down to ${\bf Z}_N$ if the
field carries $N$ units of the fundamental U(1) charge. If we
normalize the smallest nonvanishing U(1) charge as $\pm 1$, then a
VEV of U(1) charge $N$ field breaks U(1)$\rightarrow {\bf Z}_N$. So
if only even integer U(1) charge fields, including $Q=2$, are given
VEVs then the final discrete group is ${\bf Z}_2$, and we succeed in
obtaining a matter parity. However, if some phenomenological reasons
dictate some $Q=\pm 1$ fields develop GUT scale VEVs, we do not
obtain such a matter parity. In this case, we can resort only to an
approximate matter parity.

In the \Eo\ group space, the weights are divided into two classes,
the vector type \Hvec\ and the spinor type \Sp. In the U sector, the
vector type has the form of $P$ such as
\begin{equation}
P=(\pm 1~0~0~\pm 1~0~0~0),\label{VecWt}
\end{equation}
while the spinor type has the form of $P$ such as
\begin{equation}
P=\textstyle(\pm\frac12~\pm\frac12~\pm\frac12~\pm\frac12~\pm\frac12~
\pm\frac12~\pm\frac12~\pm\frac12).\label{SpWt}
\end{equation}
Below we will represent $\pm\frac12$ simply as $\pm$. In the T
sector, $P+kV_{0,\pm}$ are considered to see whether some components
are of the vector type or of the spinor type. If we pick up the
U(1)$_X$ charge of the flipped SU(5)
\begin{equation}
X=(2~2~2~2~2~0~0~0)(0^8)',\label{Xflip}
\end{equation}
the vector type weight (\ref{VecWt}) has an even number for $X$
eigenvalue while the spinor type weight (\ref{SpWt}) has an odd
number for $X$ eigenvalue. This results because we have an odd
number of entries of 2 in Eq. (\ref{Xflip}).  So this U(1)$_X$ can
be a good mother for the matter parity. If we consider three entries
of 2, it is three times U(1)$_{\rm B-L}$ of Eq. (\ref{BmL}) which
can be another good mother group for the matter parity. $Q_1, Q_2,$
and $Q_3$ can do the same job. Out of these five, only four U(1)s
are considered to be independent.

Note that all SM particles of Table \ref{tb:SM} are of \Sp\ type and
some needed Higgs doublets are of \Hvec\ type. So if all the needed
VEVs of gauge singlets are of \Hvec\ type, then we achieve
introducing an exact matter parity in that model. However, if some
\Sp\ singlet(s) is required to have a GUT scale VEV(s), then the
matter parity introduced is not exact. The charge of the
hypothetical mother U(1)$_\Gamma$ group is given in Eq.
(\ref{U1Gamma}) which has an odd number of 2 entries.

In Table \ref{tb:Singneutral}, we list gauge singlets with
U(1)$_\Gamma$ charges. Here in the $\Gamma$ column, singlets having
odd $\Gamma$ charges are boxed. We have to check whether
phenomenology needs definitely one of these to develop a large VEV
or not. If we can choose a vacuum where it is not necessary for any
of these to develop a large VEV, then we can introduce an exact
matter parity.
\begin{table}[t]
\tbl{Left-handed electromagnetically neutral $SO(10)'$ singlets.
There is only one  untwisted sector singlet $S_0$. To have a
definition of parity, $S_{15}$, $S_{16}$, $S_{18}$, $S_{20}$, and
$S_{23}$ should not develop VEVs. \label{tb:Singneutral} }
 {\begin{tabular}{|c|c|c|c|c|c|c|}
\hline  Visible states & SM notation& $B-L$ & $X$& $\Gamma$ &
$\Gamma'$&Label
\\&&&&&\\[-1.4em]
\hline $(0~0~0;0~0;1~0~-1)(0^8)'$ & ${\bf 1}_{\bf 0}(U_2)$& 0 &0 &
+2& 0&$S_0$
\\
 $(0^5;\frac{-2}{3}\frac{-2}{3}\frac{-1}{3})(\frac12\frac{-1}{2}~0;0^5)'$
 & ${\bf 1_0}(T_2^0)$& 0&0 & $+2$& 0&$S_{1}$\\
$(0^5;\frac{-2}{3}\frac{1}{3}\frac{2}{3})(\frac{-1}{2}\frac{1}{2}~0;0^5)'$
&  ${\bf 1_0}(T_2^0)$&0 &0 & $-2$& 0&$S_{2}$\\
$(0^5;\frac{1}{3}\frac{-2}{3}\frac{2}{3})(\frac{-1}{2}\frac{1}{2}~0;0^5)'$
&  ${\bf 1_0}(T_2^0)$&0 & 0& $0$& 0&$S_{3}$\\
$(0^5;\frac{1}{3}\frac{1}{3}\frac{-1}{3})(\frac{1}{2}\frac{-1}{2}~0;0^5)'$
&  $2\cdot{\bf 1_0}(T_2^0)$&0 & 0& $0$& 0&$S_{4}$\\
$(0^5;\frac{1}{3}\frac{1}{3}\frac{-1}{3})(\frac{-1}{2}\frac{1}{2}~0;0^5)'$
& $2\cdot{\bf 1_0}(T_2^0)$&0 & 0& $0$& 0&$S_{5}$\\
$(0^5;\frac{2}{3}\frac{2}{3}\frac{-2}{3})(0^8)'$ & $2\cdot{\bf
1_0}(T_4^0)$&0 &0& $0$& 0&$S_{6}$\\
$(0^5;\frac{-1}{3}\frac{-1}{3}\frac{-2}{3})(0^8)'$ & $7\cdot {\bf
1_0}(T_4^0)$&0 & 0& $+2$& 0&$S_{7}$\\
$(0^5;\frac{-1}{3}\frac{2}{3}\frac{1}{3})(0^8)'$ & $6\cdot{\bf
1_0}(T_4^0)$&0 & 0& $-2$& 0& $S_{8}$\\
$(0^5;\frac{2}{3}\frac{-1}{3}\frac{1}{3})(0^8)'$ & $6\cdot{\bf
1_0}(T_4^0)$&0 & 0& $0$& 0& $S_{9}$\\
$(0^5;1~0~0)(\frac{-1}{2}\frac{1}{2}~0;0^5)'$ &  $2\cdot{\bf
1_0}(T_6)$&0 & 0& $0$& 0&$S_{10}$\\
$(0^5;-1~0~0)(\frac{1}{2}\frac{-1}{2}~0;0^5)'$ & $2\cdot{\bf
1_0}(T_6)$&0 & 0& $0$& 0&$S_{11}$\\
 $(0^5;0~0~1)(\frac{-1}{2}\frac{1}{2}~0;0^5)'$
 & $2\cdot{\bf 1_0}(T_6)$&0 &0 &$-2$& 0&$S_{12}$\\
 $(0^5;0~0~-1)(\frac{1}{2}\frac{-1}{2}~0;0^5)'$
 & $2\cdot{\bf 1_0}(T_6)$&0 &0 & $+2$& 0&$S_{13}$\\
$(\frac{1}{4}\frac{1}{4}\frac{1}{4}\frac{1}{4}\frac{1}{4};
\frac{5}{12}\frac{5}{12}\frac{1}{12}
 )(\frac{1}{4}\frac{3}{4}~0;0^5)'$ & ${\bf 1_0}(T_1^0)$&
 $\frac12$ &$-\frac52$&
  $0$& \fbox{$+1$}&$S_{14}$\\
 $(\frac{1}{4}\frac{1}{4}\frac{1}{4}\frac{1}{4}\frac{1}{4};
\frac{5}{12}\frac{5}{12}\frac{1}{12}
 )(\frac{-3}{4}\frac{-1}{4}~0;0^5)'$ & ${\bf 1_0}(T_1^0)$&
 $\frac12$ &$-\frac52$&
  $\fbox{$-1$}$& 0&$S_{15}$\\ [0.1em]
$(\frac{-1}{4}\frac{-1}{4}\frac{-1}{4}\frac{-1}{4}\frac{-1}{4};
\frac{-1}{12}\frac{-1}{12}\frac{-5}{12}
 )(\frac{1}{4}\frac{3}{4}~0;0^5)'$ & ${\bf 1_0}(T_1^0)$&
 $-\frac12$ &$\frac52$&
 $\fbox{+1}$& 0&$S_{16}$\\
$(\frac{-1}{4}\frac{-1}{4}\frac{-1}{4}\frac{-1}{4}\frac{-1}{4};
\frac{-1}{12}\frac{-1}{12}\frac{-5}{12}
 )(\frac{-3}{4}\frac{-1}{4}~0;0^5)'$ & ${\bf 1_0}(T_1^0)$&$-\frac12$
 &$\frac52$&
 $0$& \fbox{$-1$}&$S_{17}$\\
$(\frac{1}{4}\frac{1}{4}\frac{1}{4}\frac{1}{4}\frac{1}{4};
\frac{-7}{12}\frac{5}{12}\frac{1}{12}
 )(\frac{-1}{4}\frac{-3}{4}~0;0^5)'$
& ${\bf 1_0}(T_7^0)$&$\frac12$ &$-\frac52$& $\fbox{$-1$}$ & 0
&$S_{18}$\\
$(\frac{1}{4}\frac{1}{4}\frac{1}{4}\frac{1}{4}\frac{1}{4};
\frac{-7}{12}\frac{5}{12}\frac{1}{12}
 )(\frac{3}{4}\frac{1}{4}~0;0^5)'$ & ${\bf 1_0}(T_7^0)$&$\frac12$
 &$-\frac52$&
  $0$& \fbox{$+1$}&$S_{19}$\\
$(\frac{1}{4}\frac{1}{4}\frac{1}{4}\frac{1}{4}\frac{1}{4};
\frac{5}{12}\frac{-7}{12}\frac{1}{12}
 )(\frac{-1}{4}\frac{-3}{4}~0;0^5)'$
& ${\bf 1_0}(T_7^0)$&$\frac12$ &$-\frac52$& $\fbox{+1}$& 0
&$S_{20}$\\
$(\frac{1}{4}\frac{1}{4}\frac{1}{4}\frac{1}{4}\frac{1}{4};
\frac{5}{12}\frac{-7}{12}\frac{1}{12}
 )(\frac{3}{4}\frac{1}{4}~0;0^5)'$ &
${\bf 1_0}(T_7^0)$&$\frac12$ &$-\frac52$& $+2$& \fbox{$+1$}
&$S_{21}$\\ &&&&&\\[-1.2em]
 $(\frac{-1}{4}\frac{-1}{4}\frac{-1}{4}\frac{-1}{4}\frac{-1}{4};
\frac{-1}{12}\frac{-1}{12}\frac{7}{12}
 )(\frac{-1}{4}\frac{-3}{4}~0;0^5)'$ &
${\bf 1_0}(T_7^0)$&$-\frac12$ &$\frac52$& $-2$& \fbox{$-1$}
&$S_{22}$\\
 $(\frac{-1}{4}\frac{-1}{4}\frac{-1}{4}\frac{-1}{4}\frac{-1}{4};
\frac{-1}{12}\frac{-1}{12}\frac{7}{12}
 )(\frac{3}{4}\frac{1}{4}~0;0^5)'$ &
${\bf 1_0}(T_7^0)$&$-\frac12$ &$\frac52$& $\fbox{$-1$}$& 0
&$S_{23}$\\
 $(\frac{-1}{4}\frac{-1}{4}\frac{-1}{4}\frac{-1}{4}\frac{-1}{4};
\frac{-3}{4}\frac{1}{4}\frac{1}{4}
 )(\frac{-1}{4}\frac{-3}{4}~0;0^5)'$ &
${\bf 1_0}(T_3)$&$-\frac12$ &$\frac52$& $-2$& \fbox{$-1$}
&$S_{24}$\\  &&&&&\\[-1.2em]
$(\frac{1}{4}\frac{1}{4}\frac{1}{4}\frac{1}{4}\frac{1}{4};
\frac{3}{4}\frac{-1}{4}\frac{-1}{4}
 )(\frac{1}{4}\frac{3}{4}~0;0^5)'$ &
${\bf 1_0}(T_9)$&$\frac12$ &$-\frac52$& $+2$ & \fbox{$+1$}
&$S_{25}$\\  &&&&&\\[-1.2em]
$(\frac{-1}{4}\frac{-1}{4}\frac{-1}{4}\frac{-1}{4}\frac{-1}{4};
\frac{1}{4}\frac{1}{4}\frac{-3}{4}
 )(\frac{-1}{4}\frac{-3}{4}~0;0^5)'$ & $2\cdot{\bf 1_0}(T_3)$&$-\frac12$
 &$\frac52$
& $0$& \fbox{$-1$}&$S_{26}$\\  &&&&&\\[-1.2em]
  $(\frac{1}{4}\frac{1}{4}\frac{1}{4}\frac{1}{4}\frac{1}{4};
\frac{-1}{4}\frac{-1}{4}\frac{3}{4}
 )(\frac{1}{4}\frac{3}{4}~0;0^5)'$ &
${\bf 1_0}(T_9)$& $\frac12$&$-\frac52$& $0$& \fbox{$+1$}
&$S_{27}$\\  &&&&&\\[-1.2em]
$(\frac{1}{4}\frac{1}{4}\frac{1}{4}\frac{1}{4}\frac{1}{4};
\frac{-1}{4}\frac{-1}{4}\frac{-1}{4}
 )(\frac{3}{4}\frac{1}{4}~0;0^5)'$
& $2\cdot {\bf 1_0}(T_3)$&$\frac12$ &$-\frac52$& $+2$ & \fbox{$+1$}
 &$S_{28}$ \\  &&&&&\\[-1.2em]
 $(\frac{-1}{4}\frac{-1}{4}\frac{-1}{4}\frac{-1}{4}
  \frac{-1}{4};\frac{1}{4}\frac{1}{4}\frac{1}{4})
  (\frac{-3}{4}\frac{-1}{4}~0;0^5)'$ &
$3\cdot{\bf 1_0}(T_9)$&$-\frac12$ &$\frac52$& $-2$& \fbox{$-1$}&
$S_{29}$
\\[0.2em]
\hline
\end{tabular}}
\end{table}

But to make all exotics heavy, we need $\langle S_{15}\rangle\ne 0$
and $\langle S_{23}\rangle\ne 0$. So an exact $R$-parity is not
introduced in the model. However, we note that even if we obtained
an exact $R$-parity, proton still decays by dimension-5 operators
$$
QQQL,\ u^cu^cd^ce^c.
$$
So an exact $R$-parity conservation requirement can be considered as
an over-requirement with respect to the proton longevity problem. An
approximate $R$-parity with proton lifetime comparable to that
coming from dimension-5 operators is a good enough parity. However,
if one requires an absolutely stable lightest SUSY particle (LSP),
the exact $R$-parity might be a good requirement. However, here also
the LSP with its lifetime much larger than the age of universe can
be considered as a good dark matter candidate.

The dimension-4 operator $u^cd^cd^c$ is particularly problematic if
there exists a tree level lepton number violating operator also. In
our model, $u^cd^cd^c$ carries $\Gamma=-3$, so to have a dimension-4
$u^cd^cd^c$ coupling we need composite singlets carrying $\Gamma=3$:
$\prod_i S_i$
 \begin{align}
  \underbrace{u^cd^cd^c}_{\Gamma=-3}\ \langle\underbrace{\prod_i
 S_i}_{\Gamma=+3}
 \rangle
 \end{align}
To obtain $\Gamma=+3$ from $\prod_i S_i$, we need for example
$S_0S_0S_{15}$ where $S_0$ is a $U_2$ field and $S_{15}$ is a
$T_1^0$ field. Other singlets having nonvanishing VEVs must belong
to $S_1- S_{13}$. We can check that with these combinations we
cannot satisfy the modular invariance condition $\sum {\rm
twists}=0$ mod 12. This implies that in the vacuum we choose the
coupling $u^cd^cd^c$ does not appear at all. However, the matter
parity or the $R$-parity is broken by heavy intermediate states,
which is nothing but those appearing in GUT models.  So we estimate
that the lightest neutralino lifetime is around $10^{22}$ years
through $R$-parity violating interactions, considering heavy
particles, which is long enough to be considered as a dark matter
candidate.

\section{Conclusion}

We reviewed very briefly the orbifold compactification. As a
definite acceptable example, we constructed a ${\bf Z}_{12-I}$
orbifold model with the SM gauge group and three families in which
we achieve
\begin{itemize}
\item All exotics are removed by singlet VEVs,
\item The 3rd family appears in the U sector,
\item The weak mixing angle is $\sin^2\theta_W=\frac38$, and
\item An effective $R$-parity without $u^cd^cd^c$ coupling can be introduced.
Still the neutralino can be a dark matter candidate.
\end{itemize}
In another vacuum, we can align the hypercharge so that no exotics
appear but in this case the weak mixing angle is
$\sin^2\theta_W=\frac3{14}$.

\section*{Acknowledgments}

I thank K.-S. Choi, I.-W. Kim and B. Kyae for numerous
collaborations presented in this talk. This work is supported in
part by the KRF Grants, No. R14-2003-012-01001-0 and No.
KRF-2005-084-C00001.

\end{document}